\documentclass[a4paper,10pt]{article} 

\usepackage{graphicx} 
\usepackage{amsmath,amsfonts,amssymb,bm}
\usepackage{color}

\usepackage[utf8]{inputenc}

\usepackage{microtype} 

\usepackage{ragged2e} 

\usepackage{fancyhdr}
\fancypagestyle{plain}{\pagestyle{fancy}} 

\usepackage{blindtext}

\usepackage[sort&compress,numbers]{natbib}
\usepackage{doi}

\newcommand{\dd}{{\rm d}}
\newcommand{\q}{\bm q}
\newcommand{\x}{\bm x}

\newcommand{\uu}{\bm u}
\newcommand{\vv}{\bm v}
\newcommand{\ww}{\bm w}
\newcommand{\du}{\delta_u}
\newcommand{\dv}{\delta_v}
\newcommand{\dw}{\delta_w}
\newcommand{\ubar}{\bar{\bm u}}
\newcommand{\vbar}{\bar{\bm v}}
\newcommand{\wbar}{\bar{\bm w}}
\newcommand{\cc}{\bm c}



\begin{document}

\title{Demystifying magnetic resonance measurements of the true diffusion propagator}


\author{Evren \"Ozarslan$^{a,\dagger}$, Magnus Herberthson$^b$  \\
	{}\footnotesize\it ${}^a$Department of Biomedical Engineering, Linköping University, Linköping, Sweden \\
	{}\footnotesize\it ${}^b$Department of Mathematics, Linköping University, Linköping, Sweden \\
\footnotesize\it  ${}^\dagger$E-mail: evren.ozarslan@liu.se} 
\date{}


\maketitle
\pagestyle{fancy}

\begin{abstract}
In a recent work, a method for the magnetic resonance (MR) measurement of the true diffusion propagator was introduced, which was subsequently implemented and validated for free diffusion on a benchtop MR scanner. Here, we provide a brief theoretical description of the method and discuss various experimental regimes. 
\end{abstract}

\section{Introduction}

\begin{figure}[b!]
	\begin{center}
			\includegraphics[width=.99\textwidth]{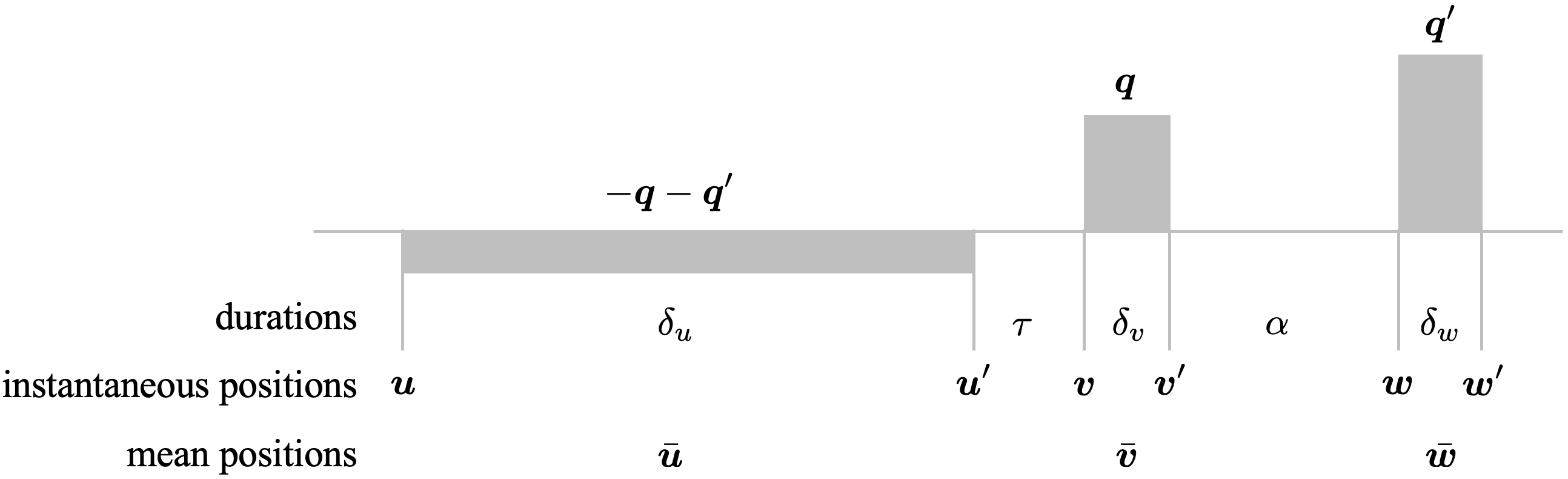}
	\end{center}
	\vspace{-14pt}\caption{Effective gradient waveform of one realization of the pulse sequence considered here along with the definitions of the employed parameters. \label{fig:definitions}}
\end{figure}

In a recent work, one of us introduced a new experimental and data analysis framework with which the true diffusion propagator can be determined through MR measurements \cite{Ozarslan21ISMRMeverything,Ozarslan21ARXIVeverything}. One realization of the gradient waveform employed by the method is illustrated in Figure \ref{fig:definitions}. This gradient waveform features three gradient pulses of durations $\du$, $\dv$, and $\dw$. The duration between the first and second pulses is $\tau$, while that between the second and third pulses is $\alpha$. The integral of the gradient vector over time multiplied by the gyromagnetic ratio determines the $q$-vectors, which are $-\q-\q'$, $\q$, and $\q'$ for the three pulses, respectively. We note that there may be many other realizations of the diffusion encoding waveform. For example, in its first experimental implementation \cite{Ordinola21ARXIVfirstexp}, each gradient pulse was replaced by a bipolar gradient pair around a 180$^\circ$ radiofrequency pulse, which made the sequence robust against complicating factors such as concomitant fields, eddy currents, and field gradients due to susceptibility variations. 

In this study, we revisit the problem and describe how this pulse sequence enables the measurement of the true diffusion propagator. Furthermore, we consider free, restricted, and general diffusion scenarios as well as another realization of the method in which the shorter pulses are applied first.

\section{General expression for the signal} 

Here, we shall employ a path integral approach for determining the MR signal intensity \cite{Yolcu15Dagstuhl}. During the application of the first pulse, the particle interrogates the positions $\uu$, $\uu_2$, $\ldots$, $\uu_{N-1}$, $\uu'$ in $d$-dimensional space. The vectors $\vv$, $\vv_2$, $\ldots$, $\vv_{N-1}$, $\vv'$ shall denote its position during the second pulse. Similarly, the particle's position vectors during the third pulse are $\ww$, $\ww_2$, $\ldots$, $\ww_{N-1}$, $\ww'$. The average position of the particle during the application of the three pulses are $\ubar=(\uu+\uu_2+\ldots+\uu')/N$, $\vbar=(\vv+\vv_2+\ldots+\vv')/N$, and $\wbar=(\ww+\ww_2+\ldots+\ww')/N$, respectively. The exact signal is obtained as $N\rightarrow\infty$, i.e., as the interval between the adjacent time points shrinks. Let $P_t(\bm a'|\bm a)$ denote the propagator indicating the differential probability for a particle initially at position $\bm a$ to be at $\bm a'$ after time $t$. In the below description, we shall drop the time subscript for propagators of infinitesimal duration for brevity. Another quantity is $\rho(\cdot)$ which is used for the probability densities that cannot be described directly by the propagator. For example, $\rho(\uu)$ is the distribution of particles in space while $\rho_{\dw}(\wbar | \ww)$ indicates the conditional probability for a particle to have the mean position $\wbar$ over a time interval of duration $\dw$ given that it is at $\ww$ at the start of the interval. Similarly, the differential probability for a random walk of duration $\du$ that starts at position $\uu$ and ends at position $\uu'$ to have the mean position $\ubar$ is denoted by $\rho_{\du}(\ubar | \uu,\uu')$. Finally, we denote by $\beta(\ubar)$ the probability density for finding a trajectory whose center of mass is located at $\ubar$---referred to as the ``bouquet density'' \cite{Ordinola21ARXIVfirstexp}.

The signal for the pulse sequence in Figure \ref{fig:definitions} can then be expressed as
\begin{align}\label{eq:E_general_clean}
	E(\q,\q') = & \int\dd\uu \int\dd\uu' \int\dd\ubar \, e^{i(\q+\q')\cdot\ubar} \int\dd\vv \int\dd\vv' \int\dd\vbar \,  e^{-i\q\cdot\vbar} \int\dd\ww \int\dd\ww' \int\dd\wbar \,  e^{-i\q'\cdot\wbar} \rho(\uu)   \nonumber\\
	& P_{\du}(\uu'|\uu) \, \rho_{\du}(\ubar | \uu,\uu') \, P_\tau(\vv|\uu') \, P_{\dv}(\vv'|\vv) \, \rho_{\dv}(\vbar | \vv,\vv') \, P_\alpha(\ww|\vv') \, P_{\dw}(\ww'|\ww) \, \rho_{\dw}(\wbar | \ww,\ww') \ ,
\end{align}
where we repeatedly made use of the following expression for the joint density for a particle originally at $\bm a$ to have the mean position $\bm{\bar a}$ over a time interval of duration $t$ and arrive to $\bm a'$ at the end of the interval:
\begin{align}\label{eq:useful}
\rho_t(\bm{\bar a},\bm a'|\bm a) & = P_t(\bm a'|\bm a) \, \rho_t(\bm{\bar a} | \bm a, \bm a') \nonumber\\
& = \lim_{N\rightarrow\infty} N^d \int\dd\bm a_2 \int\dd\bm a_3 \ldots \int\dd\bm a_{j-1} \int\dd\bm a_{j+1} \ldots \int\dd\bm a_{N-1} \nonumber\\
& \qquad\qquad P(\bm a_2|\bm a) \, P(\bm a_3|\bm a_2) \ldots P(\bm a_{j-1}|\bm a_{j-2}) \, P(N \bm{\bar a}-\bm a - \bm a_2 - \ldots -\bm a_{j-1} - \bm a_{j+1}- \ldots - \bm a'  |\bm a_{j-1}) \nonumber\\
& \qquad\qquad P(\bm a_{j+1} | N \bm{\bar a}-\bm a - \bm a_2 - \ldots -\bm a_{j-1} - \bm a_{j+1}- \ldots - \bm a' ) \, P(\bm a_{j+2}|\bm a_{j+1}) \ldots P(\bm a' |\bm a_{N-1}) \ .
\end{align}
Here we employed the change-of-variables $\bm a_j=N \bm{\bar a}-\bm a - \bm a_2 - \ldots -\bm a_{j-1} - \bm a_{j+1}- \ldots - \bm a'$ for some $j$ between 1 and $N$. 

In \eqref{eq:E_general_clean}, the integrations over $\uu$ and $\ww'$ can be tackled readily via the expressions
\begin{subequations} \label{eq:x1zN_integrals}
	\begin{align}
	\int\dd\uu \, \rho(\uu) \, P_{\delta_u}(\uu'|\uu) \, \rho_{\delta_u}(\ubar | \uu,\uu') & = \beta(\ubar)\, \rho_{\du}(\uu'|\ubar) \\
	\int\dd\ww' \, P_{\dw}(\ww'|\ww) \, \rho_{\dw}(\wbar|\ww,\ww') & = \rho_{\dw}(\wbar|\ww) \ , 
	\end{align}
\end{subequations}
yielding
\begin{align}\label{eq:gen_result}
E(\q,\q') = & \int\dd\ubar \, e^{i(\q+\q')\cdot\ubar} \int\dd\vbar \,  e^{-i\q\cdot\vbar} \int\dd\wbar \,  e^{-i\q'\cdot\wbar} \, \mathbb{I}(\ubar,\vbar,\wbar)
\end{align}
with the integrand
\begin{align}\label{eq:integrand}
\mathbb{I}(\ubar,\vbar,\wbar) = \!\int\!\dd\uu' \! \int\!\dd\vv \! \int\!\dd\vv' \! \int\!\dd\ww \, \beta(\ubar)  \, \rho_{\du}(\uu'|\ubar) \, P_\tau(\vv|\uu') \, P_{\dv}(\vv'|\vv) \, \rho_{\dv}(\vbar | \vv,\vv') \,  P_\alpha(\ww|\vv') \, \rho_{\dw}(\wbar | \ww) \, .
\end{align}

The expressions \eqref{eq:x1zN_integrals}-\eqref{eq:integrand} are quite general and can be used to evaluate the signal intensity for many processes. As a simple example, consider free diffusion. For convenience, we shall first define a Gaussian distribution, which is useful for free diffusion in $d$ dimensions with diffusivity $D$, as 
\begin{align}
G(\bm a,t)=(4\pi D t)^{-d/2} \, e^{-|\bm a|^2/4Dt} \ .
\end{align}
For free diffusion, $\rho(\bm a)=\rho$ is a constant while the other distributions needed are
\begin{subequations}
	\begin{align}\label{eq:free_quantities}
	P_t(\bm a'|\bm a) &= G(\bm a' - \bm a, t)\, , \textrm{ and} \\
	\rho_t (\bm{\bar a} | \bm a,\bm a') & = G\left(\bm{\bar a}-\frac{\bm a + \bm a'}{2}, \frac{t}{12} \right) \ ,
	\end{align}
\end{subequations}
where the first expression is the heat kernel and the second expression was derived via mathematical induction.  Employing the above expressions in Eqs. \eqref{eq:x1zN_integrals}-\eqref{eq:integrand}, one gets  
\begin{align}
E^\mathrm{free}(\q,\q') =\exp \left\{-D \left[
q^2 \left(\tau+\frac{\delta_u}{3}+\frac{\delta_v}{3}\right) +2\q\cdot\q'\left(\tau+\frac{\delta_u}{3}+\frac{\delta_v}{2}\right)+q'^2\left(\tau+\alpha+\frac{\delta_u}{3}+\delta_v+\frac{\delta_w}{3}\right) \right] \right\} \ .
\end{align}
This expression is the same as that obtained via conventional means that involves direct integration of the gradient waveform \cite{Karlicek80}.

\section{Limiting cases}

Here, we shall consider the signal in various diffusion scenarios and extreme cases of the timing parameters. These are necessary to illustrate the straightforward dependence of the observed signal to the true diffusion propagator. When the conditions are violated slightly, one will likely obtain reasonable and useful approximations to the true propagator. 

First let us take the second and third pulses to be narrow, i.e., we consider the scenario $\dv\rightarrow0$ and $\dw\rightarrow0$. In this case, the integrand \eqref{eq:integrand} takes the particularly simple form
\begin{align}\label{eq:I_vwnarrow}
\mathbb{I}(\ubar,\vbar,\wbar)  = \beta(\ubar) \left( \int\dd\uu' \rho_{\du}(\uu'|\ubar) P_\tau(\vbar|\uu') \right) P_\alpha(\wbar|\vbar) \ .
\end{align}

\subsection{The signal for restricted diffusion}

If one employs a long first pulse during which the particles can probe the entire restricted domain, the bouquet density converges to $\beta(\ubar)=\delta(\ubar-\cc)$, where $\cc$ is the center of mass of the restricted domain. Furthermore, $\rho_{\du}(\uu'|\ubar) = \rho(\uu')$, yielding
\begin{align}\label{eq:I_restricted_desired}
\mathbb{I}(\ubar,\vbar,\wbar)  = \delta(\ubar-\cc) \, A_\tau(\vbar)  \, P_\alpha(\wbar|\vbar) 
\end{align}
with 
\begin{align}
A_\tau(\vbar) = \int\dd\uu' \rho(\uu') P_\tau(\vbar|\uu') \ ,
\end{align}
where $A_\tau(\vbar)$ evaluates to $\rho(\vbar)$ for long as well as short values of $\tau$. The signal expression is obtained to be
\begin{align}\label{eq:E_restricted_desired}
E(\q,\q') = \int \dd\x \, \hat A_\tau(\x) \int \dd\x' \, \hat P_\alpha(\x'|\x) \, e^{-i(\q\cdot\x+\q'\cdot\x')}\ ,
\end{align}
where we employed the changes of variables: $\x=\vbar-\cc$ and $\x'=\wbar-\cc$. Moreover, the distributions are expressed in a frame of reference centered at the domain's center of mass, i.e., 
$\hat A_\tau(\x)=A_\tau(\x+\cc)$ and $\hat P_\alpha(\x'|\x) = P_\alpha(\x'+\cc|\x+\cc)$. 

The above expression for the case of short or long $\tau$ is employed in \cite{Ozarslan21ISMRMeverything,Ozarslan21ARXIVeverything} for the determination of the diffusion propagator in a system involving two compartments exchanging through a semipermeable membrane as well as for particles diffusing under the influence of a Hookean restoring force \cite{Uhlenbeck30}.

\subsection{The signal for free diffusion}

When diffusion is free, the bouquet density is a constant, i.e., $\beta(\ubar)=\beta$, and the most convenient experimental regime features short $\tau$ as well as short $\dv$ and $\dw$. Employing $P_\tau(\vbar|\uu')=\delta(\uu'-\vbar)$ in \eqref{eq:I_vwnarrow}, one obtains
\begin{align}\label{eq:I_free_desired}
\mathbb{I}(\ubar,\vbar,\wbar)  = \beta \, \rho_{\du}(\vbar|\ubar) \, P_\alpha(\wbar|\vbar) \ 
\end{align}
leading to the signal 
\begin{align}\label{eq:E_free_desired}
E(\q,\q')  = \int \dd\x \, \hat\rho_{\du}(\x|\bm 0) \int \dd\x' \, \hat P_\alpha(\x'|\x) \, e^{-i(\q\cdot\x+\q'\cdot\x')}\ ,
\end{align}
where we employed the changes of variables $\x=\vbar-\ubar$ and $\x'=\wbar-\ubar$ and exploited the translational invariance of the distributions involved.

The above expression is employed in \cite{Ordinola21ARXIVfirstexp} for measuring the diffusion propagator for freely diffusing molecules.

\subsection{From the signal to the diffusion propagator}

The form of the signal in \eqref{eq:E_restricted_desired} and \eqref{eq:E_free_desired} makes determining the diffusion propagator possible. In either case, a special case of the experiment ($\q'=\bm 0$) is incorporated and inverse Fourier transforms are employed. Specifically, this is accomplished through the expression \cite{Ozarslan21ISMRMeverything,Ozarslan21ARXIVeverything}
\begin{align} \label{eq:P_from_E}
\hat P_\alpha(\x'|\x) & =\frac{\int\dd\q\, e^{i\q\cdot\x}\int\dd\q'\, e^{i\q'\cdot\x'}\, E(\q,\q')}{(2\pi)^d \int\dd\q\,e^{i\q\cdot\x}\, E(\q,\bm 0)} \ .
\end{align}

\subsection{The signal for other diffusion scenarios}

One can once again employ, e.g., the regime in which $\dv$, $\dw$, and $\tau$ are short so that 
\begin{align}\label{eq:I_other_desired}
\mathbb{I}(\ubar,\vbar,\wbar)  = \beta(\ubar) \, \rho_{\du}(\vbar|\ubar) \, P_\alpha(\wbar|\vbar) \ ,
\end{align}
and the signal is 
\begin{align}\label{eq:E_other_desired}
E(\q,\q')  = \int \dd\x \int \dd\x' \, e^{-i(\q\cdot\x+\q'\cdot\x')} \int \dd\ubar \, \beta(\ubar) \,  \rho_{\du}(\ubar+\x|\ubar) \, P_\alpha(\ubar+\x'|\ubar+\x) \ ,
\end{align}
where we employed the change of variables $\x=\vbar-\ubar$ and $\x'=\wbar-\ubar$ as before.

Employing \eqref{eq:P_from_E} yields an ``apparent'' propagator
\begin{align} 
P_\alpha^\mathrm{app}(\x'|\x) & =\frac{\int \dd\ubar \, \beta(\ubar) \,  \rho_{\du}(\ubar+\x|\ubar) \, P_\alpha(\ubar+\x'|\ubar+\x)}{(2\pi)^d \int \dd\ubar \, \beta(\ubar) \,  \rho_{\du}(\ubar+\x|\ubar) } \ .
\end{align}
Depending on the medium being studied, further assumptions and simplifications can be made for the distributions involved.

\subsection{Another realization of the technique}

\begin{figure}[b!]
	\begin{center}
		\includegraphics[width=.99\textwidth]{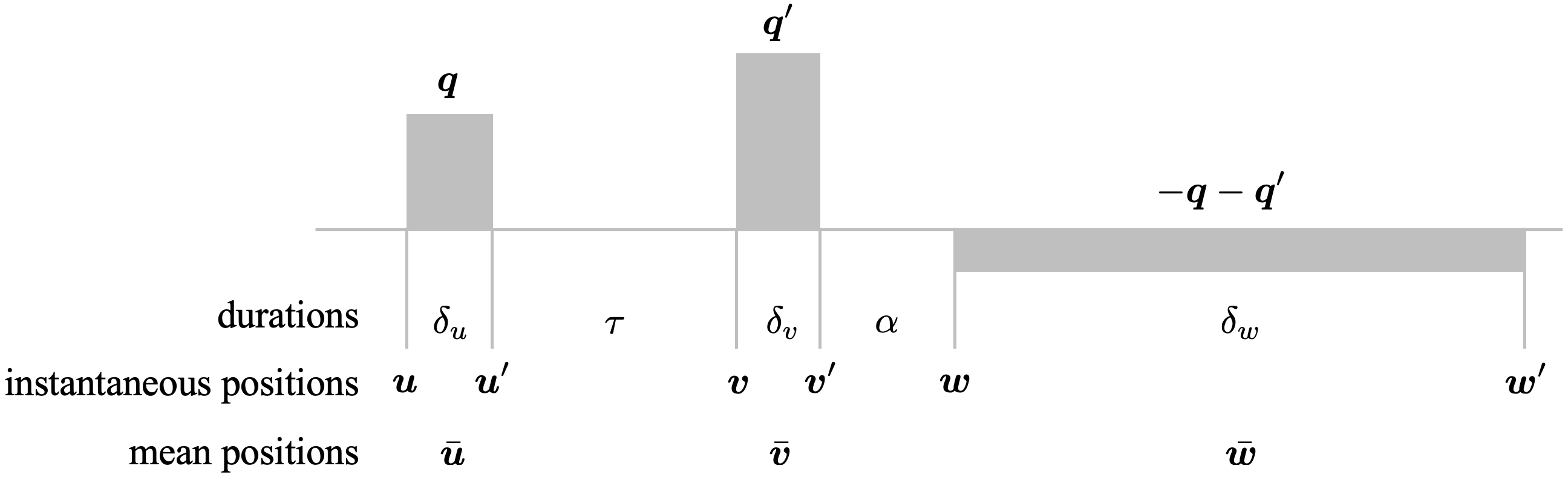}
	\end{center}
	\vspace{-14pt}\caption{Effective gradient waveform of another realization of the technique featuring two narrower pulses followed by the third pulse. \label{fig:definitions_ssl}}
\end{figure}

Here, we stick to the three-pulse sequence, but consider the case in which the first two pulses are narrow. The relevant parameters are provided  in Figure \ref{fig:definitions_ssl}. 

For restricted diffusion, when the third pulse is long, the integrand becomes 
\begin{align}\label{eq:I_ssl_rest}
\mathbb{I}(\ubar,\vbar,\wbar)  = \rho(\ubar) \, \delta(\wbar - \bm c) \, P_\tau(\vbar|\ubar) \ .
\end{align}
Thus, the $2d$-dimensional inverse Fourier transform of the signal is $\hat\beta(\x)\, \hat P_\tau(\x'|\x)$, where $\x=\ubar-\bm c$ and $\x'=\vbar-\bm c$ and the distributions are expressed in a frame of reference whose origin is at the center of mass of the restricted domain. The $d$-dimensional inverse Fourier transform of $E(\q,\bm 0)$ yields $\hat\beta(\x)$. Thus, dividing the two results as in \eqref{eq:P_from_E} reveals the propagator  $\hat P_\tau(\x'|\x)$. 

The same procedure yields the propagator for free diffusion as well when $\du$, $\dv$, and $\alpha$ are all short. In this case, and for general media, the following apparent propagator is obtained:
\begin{align} \label{eq:P_from_E_ssl}
P_\tau^\mathrm{app}(\x'|\x) & =\frac{\int \dd\wbar \, \beta(\wbar+\x) \,  P_\tau(\wbar+\x'|\wbar+\x) \, \rho_{\dw}(\wbar|\wbar+\x')}{(2\pi)^d \int \dd\wbar \, \int \dd\x' \, P_\tau(\wbar+\x'|\wbar+\x) \, \rho_{\dw}(\wbar|\wbar+\x')} \ .
\end{align}

\section{Discussion and Conclusion}

In this work, we demystified a recently introduced framework \cite{Ozarslan21ISMRMeverything,Ozarslan21ARXIVeverything,Ordinola21ARXIVfirstexp} for determining the true diffusion propagator via magnetic resonance. We discussed various regimes in which the method yields exact results. Remarkably, the procedure for reconstructing the propagator is valid for free diffusion and fully restricted domains. In the presence of deviation from the ideal experimental regimes and these diffusion scenarios, the application of the framework would reveal an apparent diffusion propagator, which could still provide new insights into the diffusive dynamics and the ambient environment that influences it.

\end{document}